\documentclass[epj]{svjour}

\usepackage[autostyle,italian=guillemets]{csquotes}
\usepackage[utf8]{inputenc}
\usepackage[nowrite,infront,standard]{frontespizio}
\usepackage[english]{babel}
\usepackage{physics}
\usepackage{amsmath,amssymb}
\usepackage [svgnames]{xcolor}
\usepackage{xcolor}
\newcommand{\new}[1]{{\color{black} #1}}
\usepackage{mathrsfs}
\usepackage{subfloat}
\usepackage{frontespizio}
\usepackage{chemmacros}
\usepackage{float}
\usepackage{graphicx} 
\usepackage{bm}
\usepackage{appendix}
\usepackage{hyperref}
\usepackage{midpage}
\usepackage{placeins}
\usepackage{subfig}

\begin{document}
\title{Complete complementarity relations for quantum correlations in neutrino oscillations}
\author{ V.~A.~S.~V.~Bittencourt \inst{1}, M.~Blasone\inst{2,3}, S.~De Siena\inst{4} and C.~Matrella\inst{2,3}
}                     
%
%
\institute{Max Planck Institute for the Science of Light, 91058 Erlangen, Germany \and Dipartimento di Fisica, Universit\`a degli Studi di Salerno, Via Giovanni Paolo II, 132 84084 Fisciano, Italy \and INFN, Sezione di Napoli, Gruppo Collegato di Salerno, Italy \and Universit\`a degli Studi di Salerno (Ret.),  email: silvio.desiena@gmail.com }
\date{Received: date / Revised version: date}
%
\abstract{We analyze quantum correlations and quantum coherence in neutrino oscillations. To this end, we exploit complete complementarity relations (CCR) that fully characterize the interplay between different correlations encoded in a quantum system both for pure and mixed states. We consider the CCR for neutrino oscillations both in the case of plane-waves (pure state) and of wave packets (mixed state). In this last case we find a complex structure of correlations depending on the mixing angle, and we show the connection with the non local advantage of quantum coherence, a relevant quantifier of coherence.
\PACS{
      {PACS-key}{discribing text of that key}   \and
      {PACS-key}{discribing text of that key}
     } 
} 
\authorrunning{V. Bittencourt, M.~Blasone, S.~De~Siena and C.~Matrella}
\maketitle
\section{Introduction}
\label{intro}

The study of quantum properties intrinsic to neutrino oscillations, has recently attracted much attention \cite{Blas0}-\cite{Li}. This is motivated by the possibility of exploiting  neutrinos as a resource for quantum information tasks, with distinct characteristics with respect to photons. On the other hand, such studies help in catching the fundamental properties of these elementary particles which represent an ideal system for testing various aspects of quantum correlations.

In the last few years, investigation of quantum correlations and quantum coherence has led to considerable advancement and understanding of these concepts \cite{XiLi,Adesso:2016ygq,Streltsov:2016iow}. In particular, complete complementarity relations (CCR) \cite{Greenberger}-\cite{Englert} allow the full characterization of  different  correlations encoded in a quantum system, 
 both for pure and mixed states. The concept of complementarity has been originally formulated in the context of two-slit experiment, where one defines a predictability, associated to the knowledge of the path of the particle, and the visibility, connected to the capacity of distinguishing interference fringes \cite{Wootters}. 
By denoting the predictability as $P$ and the visibility as $V$, the  complementarity relation takes the form:
\begin{equation}
P^{2}+V^{2}\le 1.
\label{i1}
\end{equation}
This equation is saturated only for pure states. For mixed states the strict inequality holds. In Ref.\cite{Jakob2010} it is shown that \new{the lack of knowledge implied by the strict inequality} is due to the presence of entanglement between two subsystems. However, as pointed out in Ref.\cite{Basso2020}, entanglement is not the only quantum correlation existing in multipartite systems and consequently a modification of the predictability is required in order to obtain a CCR for mixed states. 

The quantum nature of neutrino oscillations has been investigated \new{in terms of} entanglement \cite{Blas0,Blas2,Banerjee:2015mha,Alok}, Bell and Leggett-Garg inequalities\cite{Gangopadhyay:2017nsn,formaggio,Naikoo2,Ma}, and various aspects of quantum coherence including steering \cite{Ming}, NAQC \cite{Ming,Wang,Us}, \new{and} entropic uncertainty relations \cite{Wang}. It would be desirable to find the connections among such many different facets of quantumness in such a system, \new{thus unifying the description of quantum correlations in neutrino oscillations.}

In this Letter we consider the CCR for neutrino oscillations, both in the plane-wave case and in the wave-packet approach. In the first instance, we deal with a pure state, while for the case of wave-packets, one has a mixed state leading to a complex internal structure of correlations. We achieve a complete characterization of correlations occurring in the system of oscillating neutrinos. We find that some of these correlations persist also when oscillations are washed out, i.e. in the long-distance limit. We are also able to recognize various quantities already studied in the literature, within the different terms of CCR.

In the following, we first briefly introduce CCR for pure and mixed states and then apply these relations to the case of neutrinos both in the plane wave description and in the wave packet approach. We consider data for three different experiments, Daya Bay, MINOS and KAMLAND. Mathematical details are included as Supplementary Material in a separate file.

\section{CCR for pure and mixed states}
\label{sec:1}

In this section, we introduce the CCR in a bipartite state represented as a vector in the Hilbert state $\mathcal{H}_{A}\otimes \mathcal{H}_{B}$ of dimension $d=d_{A}d_{B}$, where $d_{A}$ and $d_{B}$ are the dimension of the subsystem A and B, respectively. If we label with $\{\ket{i}_{A}\}_{i=0}^{d_{A}-1}$ and $\{\ket{j}_{B}\}_{j=0}^{d_{B}-1}$ the local basis for the spaces $\mathcal{H}_{A}$ and $\mathcal{H}_{B}$,  $\{\ket{i}_{A}\otimes \ket{j}_{B}=\ket{i,j}_{AB}\}_{i,j=0}^{d_{A}-1,d_{B}-1}$ represents an orthonormal basis for $\mathcal{H}_{A}\otimes \mathcal{H}_{B}$. In this basis, the density matrix of any bipartite state is given by
\begin{equation}
\rho_{A,B}=\sum_{i,k=0}^{d_{A}-1} \sum_{j,l=0}^{d_{B}-1}\rho_{ij,kl}\ket{i,j}\bra{k,l}.
\label{0.2}
\end{equation}
For pure states $\rho_{ij,kl}=a_{ij}a_{kl}^{*}$. The state of subsystem A(B) is obtained by tracing over B(A). For example, for subsystem A, we have:
\begin{equation}
\rho_{A}=\sum_{i,k=0}^{d_{A}-1} \left( \sum_{j=0}^{d_{B}-1} \rho_{ij,kj} \right)\ket{i}_{A}\bra{k} \equiv \sum_{i,k=0}^{d_{A}-1} \rho_{ik}^{A}\ket{i}_{A}\bra{k},
\label{0.3}
\end{equation}
with a similar form for the subsystem B.

In general, \new{even if the joint state $\rho_{A,B}$ is pure,} the states of the subsystems A and B are not pure, which implies that \new{some information is missing when the state of a single subsystem is considered.} If the state of the subsystem A is mixed then $1-Tr\rho_{A}^{2}>0$, which yields a  complementarity relation \cite{Basso2020}:
\begin{equation}
P_{{\rm{hs}}}(\rho_{A})+C_{\rm{hs}}(\rho_{A})<\frac{d_{A}-1}{d_{A}}
\label{0.4}
\end{equation}
where $P_{\rm{hs}}(\rho_{A})=\sum_{i=0}^{d_{A}-1}(\rho_{ii}^{A})^{2}-\frac{1}{d_{A}}$ is the predictability measure and $C_{\rm{hs}}(\rho_{A})=\sum_{i\ne k}^{d_{A}-1}|\rho_{ik}^{A}|^{2}$ is the Hilbert-Schmidt quantum coherence \cite{Maziero}. The information content absent from system A is represented by the strict inequality. The missing information is being shared via correlations with the subsystem B \cite{Qian}. In fact, for pure states the unity of the density matrix's trace $1-\rm{Tr}\rho_{A,B}^{2}=0$ can be written in the form of the complete complementarity relation:
\begin{equation}
P_{\rm{hs}}(\rho_{A})+C_{\rm{hs}}(\rho_{A})+C_{\rm{hs}}^{nl}(\rho_{A|B})=\frac{d_{A}-1}{d_{A}}
\label{0.5}
\end{equation}
where $C_{\rm{hs}}^{nl}(\rho_{A|B})=\sum_{i\ne k, j\ne l}|\rho_{ij,kl}|^{2}-2\sum_{i\ne k, j<l}\Re (\rho_{ij,kj} \rho_{il,kl}^{*})$ is called non local quantum coherence, that is the coherence shared \new{between A and B.} 

Another form of CCR can be obtained \new{by defining the predictability and the coherence measures in terms of the von Neumann entropy. In this case, the CCR reads \cite{B20}
\begin{equation}
C_{{\rm{re}}}(\rho_{A})+P_{{\rm{vn}}}(\rho_{A})+S_{{\rm{vn}}}(\rho_{A})=\log_2 d_{A}.
\label{12}
\end{equation}
Here $C_{{\rm{re}}}(\rho_{A})=S_{{\rm{vn}}}(\rho_{A,\, {\rm{diag}}})-S_{{\rm{vn}}}(\rho_{A})$ is the relative entropy of coherence, with $S_{{\rm{vn}}}(\rho)$ denoting the von Neumann entropy of  $\rho$, and $\rho_{A,\, {\rm{diag}}}=\sum_{i=1}^{d_{A}} \rho_{ii}^{A}\ket{i}\bra{i}$. $P_{{\rm{vn}}}(\rho_{A})\equiv\log_2 d_{A}-S_{{\rm{vn}}}(\rho_{A,\, {\rm{diag}}})$, is a measure of predictability. 
For pure states $S_{{\rm{vn}}}(\rho_{A})$ is a measure of entanglement between A and B.}

We now turn our attention to the case of a neutrino state,
\begin{equation}
\ket{\nu_{\alpha}(t)}=a_{\alpha\alpha}(t)\ket{\nu_{\alpha}}+a_{\alpha\beta}(t)\ket{\nu_{\beta}},
\label{n1}
\end{equation}
where $\alpha(\beta)$ denote flavors. We can then using the following correspondence \cite{Blas0,Blas2}
\begin{equation}
\begin{split}
\ket{\nu_{\alpha}}&=\ket{1}_{\alpha} \otimes \ket{0}_{\beta}=\ket{10},\\
\ket{\nu_{\beta}}&=\ket{0}_{\alpha} \otimes \ket{1}_{\beta}=\ket{01},
\end{split}
\label{n2}
\end{equation}
\new{where it is highlighted the composite nature of neutrino flavor states.} Eq. (\ref{n1}) for an initial electronic neutrino becomes:
\begin{equation}
\ket{\nu_{e}(t)}=a_{ee}(t)\ket{10}+a_{e\mu}(t)\ket{01},
\label{n3}
\end{equation}
and the density matrix, in the basis $\{\ket{00},\ket{01},\ket{10},\ket{11}\}$, reads:
\begin{equation}
\begin{split}
\rho_{A,B}=
\begin{pmatrix}
0&0&0&0\\
0&|a_{e\mu}(t)|^{2}&a_{ee}(t)a^{*}_{e\mu}(t)&0\\
0&a_{e\mu}(t)a^{*}_{ee}(t)&|a_{ee}(t)|^{2}&0\\
0&0&0&0
\end{pmatrix}.
\end{split}
\label{n4}
\end{equation}
The state of subsystems A and B are:
\begin{equation}
\rho_{A}=\begin{pmatrix}
|a_{ee}(t)|^{2}&0\\
0&|a_{e\mu}(t)|^{2}
\end{pmatrix}
 \quad, \qquad
\rho_{B}=\begin{pmatrix}
|a_{e\mu}(t)|^{2}&0\\
0&|a_{ee}(t)|^{2}
\end{pmatrix}
\label{n6}
\end{equation}

\new{From Eqs. (\ref{n4}),(\ref{n6}) we find that $P_{\rm{hs}}(\rho_{A})=P_{ee}^{2}+P_{e\mu}^{2}-\frac{1}{2}$, $C_{\rm{hs}}(\rho_{A})=0$ and $C_{\rm{hs}}^{nl}(\rho_{AB})=2P_{ee}P_{e\mu}$, where we use $|a_{ee}(t)|^{2}=P_{ee}$, $|a_{e\mu}(t)|^{2}=P_{e\mu}$ and $P_{ee}+P_{e\mu}=1$. Since the state \eqref{n4} is pure, Eq.(\ref{0.5}) is verified.} Furthermore, considering Eq.(\ref{n6}) is simple to see that $\rho_{A}=\rho_{A,\, {\rm{diag}}}$ and, consequently, $S_{{\rm{vn}}}(\rho_{A})=S_{{\rm{vn}}}(\rho_{A,\, {\rm{diag}}})$. As result, $C_{{\rm{re}}}(\rho_{A})=0$, $P_{{\rm{vn}}}(\rho_{A})=|a_{ee}|^{2}\log_{2}|a_{ee}|^{2}+|a_{e\mu}|^{2}\log_{2}|a_{e\mu}|^{2}$ and $S_{{\rm{vn}}}(\rho_{A})=-|a_{ee}|^{2}\log_{2}|a_{ee}|^{2}-|a_{e\mu}|^{2}\log_{2}|a_{e\mu}|^{2}$. Since the dimension of subsystem A is $d_{A}=1$ then $\log_{2}d_{A}=0$ and Eq.(\ref{12}) is satisfied.\\

\new{The CCRs \eqref{0.5} and \eqref{12} are valid only for a pure density matrix $\rho_{AB}$. For mixed states, the CCR have to be modified to correctly quantify the complementarity behaviour of subsystem A \cite{B21},\cite{Angelo}. 
For instance, in this case $S_{{\rm{vn}}}(\rho_{A})$ cannot be considered as a measure of entanglement, but it is just a measure of mixedness of A. For mixed states, the correct CCR is given by}
\begin{equation}
\log_{2} d_{A}=I_{A:B}(\rho_{AB}) +S_{A|B}(\rho_{AB})+P_{{\rm{vn}}}(\rho_{A})+C_{re}(\rho_{A}),
\label{13}
\end{equation}
where $I_{A:B}(\rho_{AB}) $ is the mutual information of A and B and $S_{A|B}(\rho_{AB})=S_{{\rm{vn}}}(\rho_{AB})-S_{{\rm{vn}}}(\rho_{B})$. This CCR constrains the local aspects of A by its correlations with B, given by $I_{A:B}(\rho_{AB})$ and the remaining ignorance about A given that we have access to the system B. In other words, we can consider $ S_{A|B}(\rho_{AB})$  as a quantity that measures the ignorance about the whole system that we have by looking only to subsystem A.

\section{Correlations in neutrino wave-packets with mixed-state CCR}

In principle, a neutrino system is described by a pure state\new{, such as the one in Eq.(\ref{n1})}. By using a wave packet approach \cite{Giunti},\cite{Giunti2}, the density matrix describing the evolution of a neutrino state $\rho_{\alpha}(x,t)$ depends on both the position and time. Typically, such density matrix is integrated over time, which yields \cite{Blas1}
\begin{equation}
\rho_{\alpha}(x)=\sum_{k,j}U_{\alpha k}U_{\alpha j}^{*} f_{jk}(x) \ket{\nu_{j}}\bra{\nu_{x}},
\label{14}
\end{equation}
where $ f_{jk}(x)= \exp\biggl[-i \frac{\Delta m_{jk}^{2}x}{2E}-\biggl(\frac{\Delta m_{jk}^{2}x}{4 \sqrt{2} E^{2} \sigma_{x}}\biggl)^{2}\biggl]$.  We express $\rho_{\alpha}(x)$ in terms of flavor eigenstates by establishing the identification $\ket{\nu_{\alpha}}=\ket{\delta_{\alpha e}}_{e}\ket{\delta_{\alpha \mu}}_{\mu}\ket{\delta_{\alpha \tau}}_{\tau}$. By using the relation $\ket{\nu_{i}}=\sum_{\alpha}U_{\alpha i}\ket{\nu_{\alpha}}$, we can write:
\begin{equation}
\rho_{\alpha}(x)=\sum_{\beta\gamma}F^{\alpha}_{\beta\gamma}(x)\ket{\delta_{\beta e}\delta_{\beta \mu}\delta_{\beta \tau}}\bra{\delta_{\gamma e}\delta_{\gamma \mu}\delta_{\gamma\tau}}
\label{15}
\end{equation}
where
\begin{equation}
F^{\alpha}_{\beta\gamma}(x)=\sum_{kj}U_{\alpha j}^{*}U_{\alpha k} f_{jk}(x) U_{\beta j} U_{\gamma k}^{*}
\label{16}
\end{equation}
The density matrix \eqref{14} represents a mixed state, and consequently to understand the interplay between the different quantum correlations encoded in the state, one has to consider the CCR given by Eq.(\ref{13}) (see  Supplementary material). In the following, we show \new{results for} two-flavor neutrino mixing and a state that is initially an electron neutrino, such that:
\begin{equation}
\rho_{e\mu}(x)=\begin{pmatrix}
0&0&0&0\\
0&F^{e}_{ee}(x)&F^{e}_{e\mu}(x)&0\\
0&F^{e}_{\mu e}(x)& F^{e}_{\mu\mu}(x)&0\\
0&0&0&0
\end{pmatrix}
\label{8}
\end{equation}
The  right hand side terms of Eq.(\ref{13}) can then be evaluated, with the explicit expressions given in the supplementary material. \new{In particular, the local coherence of the subsystem $\rho_{e}$ (the last term of Eq.\eqref{13}) vanishes for the state \eqref{8}} indicating that the two subsystems do not display an internal structure.

We recognize in the non-local terms of the CCR several quantum correlations already studied in literature. We find that the sum of the first two terms of Eq.(\ref{13}) is equal to the Quantum Discord,  a measure of nonclassical correlations between two subsystems of a quantum system, defined as \cite{Ollivier}:
\begin{equation}
QD(\rho_{AB})=I(\rho_{AB})-CC(\rho_{AB}),
\label{20}
\end{equation}
where $I(\rho_{AB})=S_{\rm{vn}}(\rho_{A})+S_{\rm{vn}}(\rho_{B})-S_{\rm{vn}}(\rho_{AB})$ represent the total correlations between the subsystems A and B and $CC(\rho_{AB})= \max_{\{\Pi_{i}^{b}\}}\biggl(S_{\rm{vn}}(\rho_{B})-S_{{\rm{vn}},\{\Pi_{i}^{b}\}}(\rho_{A|B})\biggl)$, quantifies the classical correlations obtained with a maximization over the set of all possible positive operator-valued measures $\{\Pi_{i}^{b}\}$ on the subsystem B with outcomes $b=\{0,1\}$. Hence, the quantum part is:
\begin{equation}
QD(\rho_{AB})=S_{\rm{vn}}(\rho_{B})-S_{\rm{vn}}(\rho_{AB})+\min_{\{\Pi_{i}^{b}\}} S_{{\rm{vn}},\{\Pi_{i}^{b}\}}(\rho_{A|B}).
\label{21}
\end{equation}
The evaluation of the Quantum Discord for the density matrix in Eq. (\ref{8}) gives us:

\begin{equation}
QD(\rho_{e\mu})=-F^{e}_{ee}\log_{2}F^{e}_{ee}-F^{e}_{\mu\mu}\log_{2}F^{e}_{\mu\mu},
\label{22}
\end{equation}
which corresponds precisely to the sum of the first two terms of the right hand side of Eq.(\ref{13}) (see Supplementary material).

It is also interesting to investigate the connection existing with the Non-Local Advantage of Quantum Coherence \new{(NAQC) \cite{Mondal}. Such quantum correlation occurs in a bipartite system when the average coherence of the conditional state of a subsystem B, after a local measurements on A, exceeds the coherence limit of the single subsystem. In the hierarchy of quantum correlations, NAQC has been classified as the strongest, overaking the Bell non-locality \cite{HuWang}.} Several definitions of NAQC have been formulated which differ in the distinct coherence measures used. Here, we consider that based on the relative entropy of coherence \new{(see equation \eqref{12})}.
Given a state $\rho$ in the reference basis $\{\ket{i}\}$, a measure of coherence takes the form $C_{D}(\rho)=\min_{\delta\in I}D(\rho,\delta)$, 
that is the minimum distance between $\rho$ and the set of incoherent states $I$. $D(\rho,\delta)$ is a distance measure between two quantum states. For example, one can consider  $D(\rho,\delta)=S(\rho||\delta)$, the quantum relative entropy. By minimizing over the set of incoherent states, one can obtain two bona fide measures of coherence \cite{Braumgratz} as:
\begin{equation}
 C_{{\rm{re}}}(\rho)=S_{\rm{vn}}(\rho_{{\rm{diag}}})-S_{\rm{vn}}(\rho),
\label{24}
\end{equation}
where $S_{\rm{vn}}(\rho)$ is the von Neumann entropy of $\rho$ and $\rho_{{\rm{diag}}}$ is the matrix of the diagonal elements of $\rho$.\\

Mondal et al. \cite{Mondal} defined the NAQC of a bipartite state $\rho_{AB}$ considering the average coherence of the post measurement state $\{ p_{B|\Pi_{i}^{a}},\rho_{B|\Pi_{i}^{a}}\}$ of B after a local measurement $\Pi_{i}^{a}$ on A:
\begin{equation}
N(\rho_{AB})=\frac{1}{2}\sum_{i\ne j, a =\pm} p_{B|\Pi_{i}^{a}}C^{\sigma_{j}}(\rho_{B|\Pi_{i}^{a}}),
\label{25}
\end{equation}
where $\Pi_{i}^{\pm}=\frac{I\pm\sigma_{i}}{2}$, with $I$ and $\sigma_{i},(i=1,2,3)$ being the identity and the three Pauli operators; $ p_{B|\Pi_{i}^{a}}=\Tr(\Pi_{i}^{a} \rho_{AB})$, $\rho_{B|\Pi_{i}^{a}}=\Tr_{A}(\Pi_{i}^{a} \rho_{AB})/ p_{B|\Pi_{i}^{a}}$. $C^{\sigma_{j}}(\rho_{B|\Pi_{i}^{a}})$ is the coherence of the conditional state of B with respect to the eigenbasis of $\sigma_{j}$. 

Evaluating Eq.(\ref{25}) for the state Eq.(\ref{8}), we find (see Supplementary material):
\begin{equation}
N(\rho_{e\mu})=2-F^{e}_{ee}\log_{2}F^{e}_{ee}-F^{e}_{\mu\mu}\log_{2}F^{e}_{\mu\mu},
\label{26}
\end{equation}
and it is immediate to find the relation $N(\rho_{e\mu})=2+I_{e:\mu}(\rho_{e\mu})+S_{e|\mu}(\rho_{e\mu})$, i.e:
\begin{equation}
 N(\rho_{e\mu})=2+QD(\rho_{e\mu}).
\label{27}
\end{equation}
\\

Let us now analyze the neutrino oscillation \new{in the light of} CCR, by using the parameters (see Table) from the Daya Bay \cite{Daya},\cite{Daya1}, KamLAND \cite{Kam},\cite{Kam1} and MINOS \cite{Minos},\cite{Minos1} experiments. Daya-Bay and KamLAND are electron-antineutrino disappearance experiments, while MINOS is a muonic neutrino disappearance experiment.
\begin{center}
    \begin{tabular}{|  p{4.5cm} |  p{4.5cm} | p{4.5cm} |}
    \hline
   Daya-Bay & KamLAND  & MINOS \\ \hline
   $\Delta m_{ee}^{2}=2.42^{+0.10}_{-0.11}\times 10^{-3} eV^{2}$ &   $\Delta m_{12}^{2}=7.49\times 10^{-5}$ & $\Delta m_{32}^{2}=2.32^{+0.12}_{-0.08}\times 10^{-3} eV^{2}$\\ \hline
 $\sin^{2} 2\theta_{13} = 0.084^{+0.005}_{-0.005}$ & $\tan^{2} 2 \theta_{12}=0.47$  & $\sin^{2} 2\theta_{23} = 0.95^{+0.035}_{-0.036}$ \\ \hline
$L\in[364 m,1912 m]$ & $L=180$ Km & $L=735$ km\\
    \hline
$E\in[1 MeV, 8 MeV]$ & $E\in[2 MeV, 10 MeV]$ & $E\in[0.5 GeV, 50 GeV]$ \\ \hline
    \end{tabular}
\label{table1}
\end{center}

\begin{figure}[t]
\centering
 \subfloat[][\emph{DAYA BAY }]{{\includegraphics[width =6 cm]{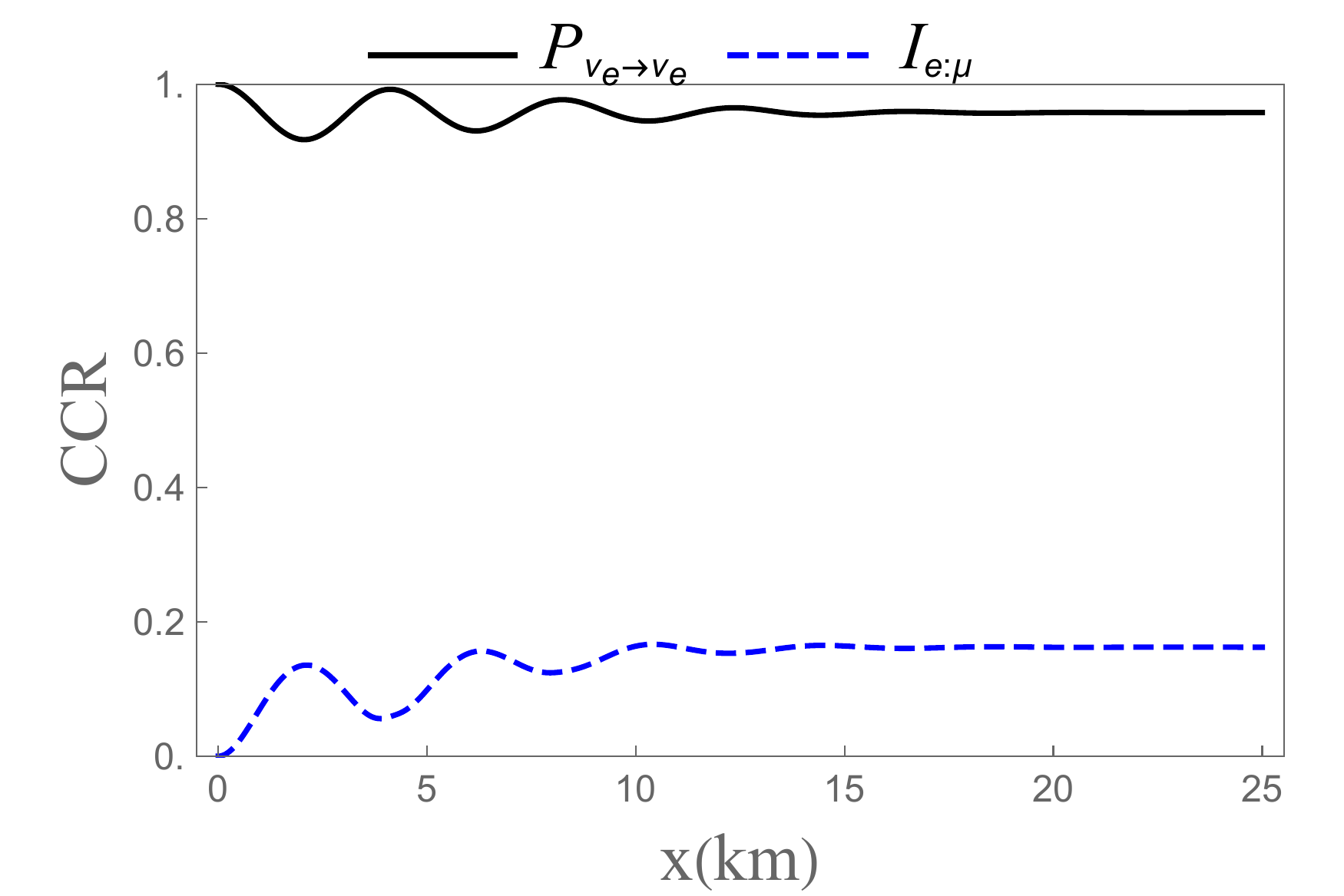}}\quad\quad
{\includegraphics[width =6.4 cm]{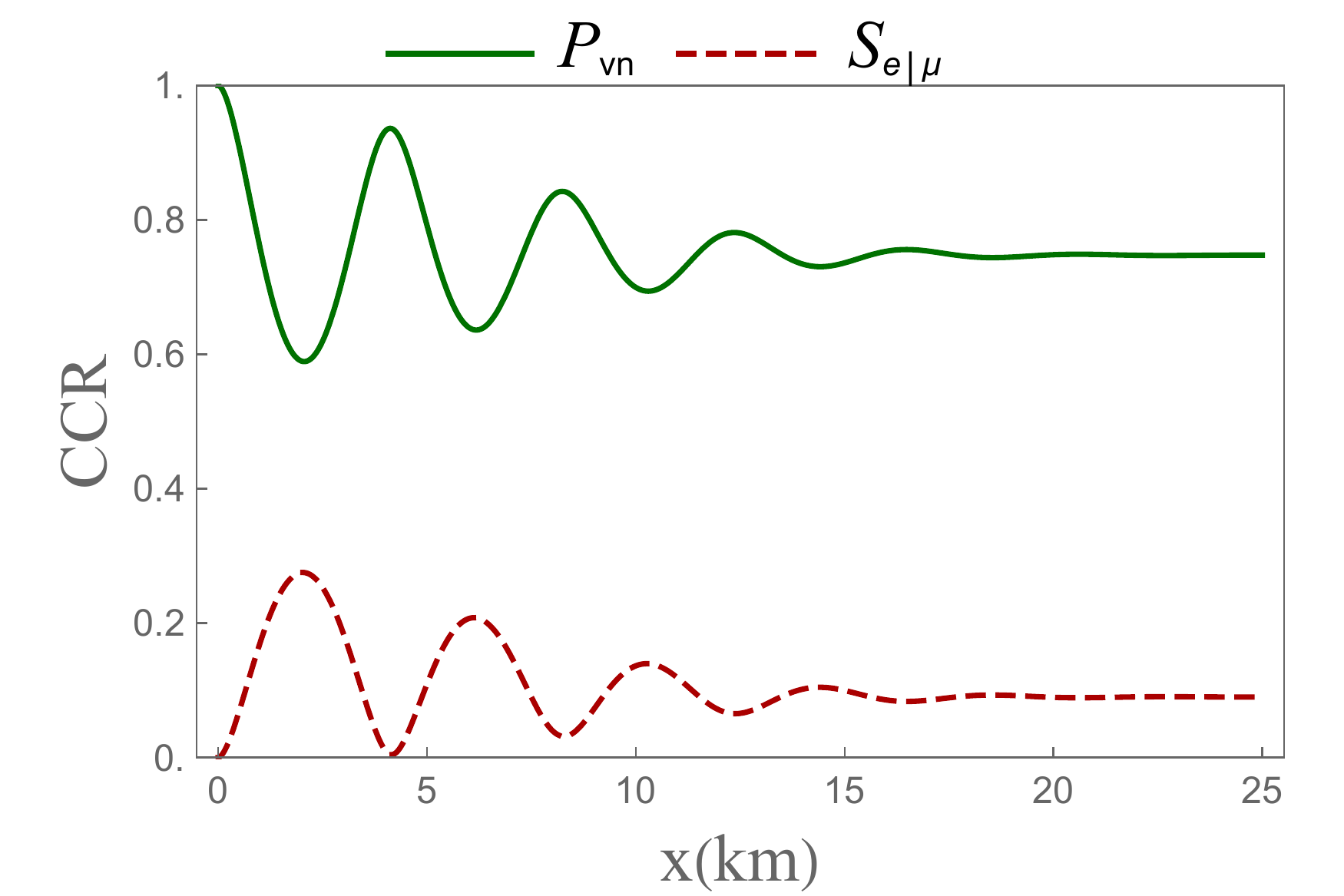}}}\\
 \subfloat[][\emph{KamLAND }]{{\includegraphics[width =6 cm]{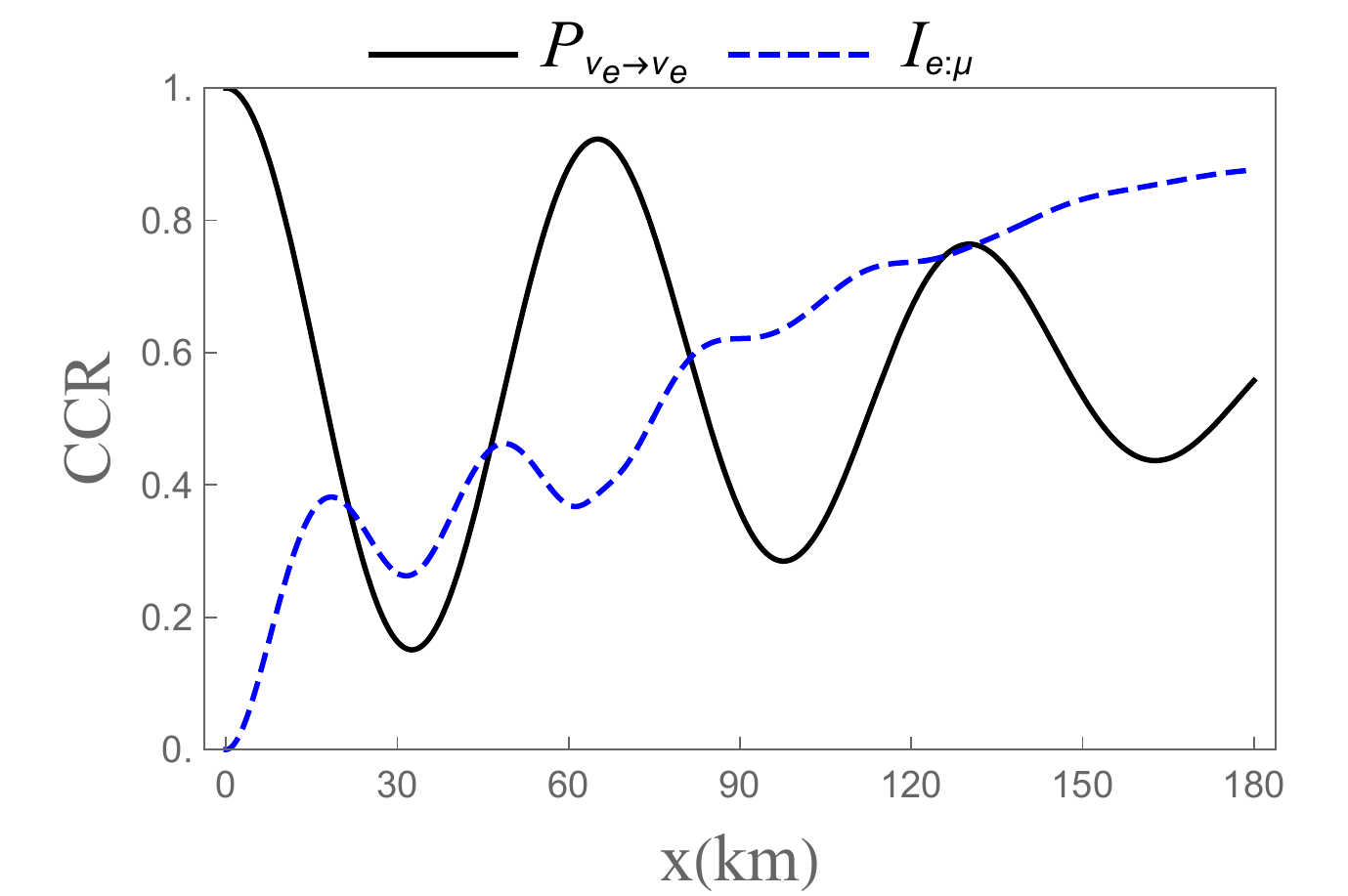}}\quad\quad
{\includegraphics[width =6.4 cm]{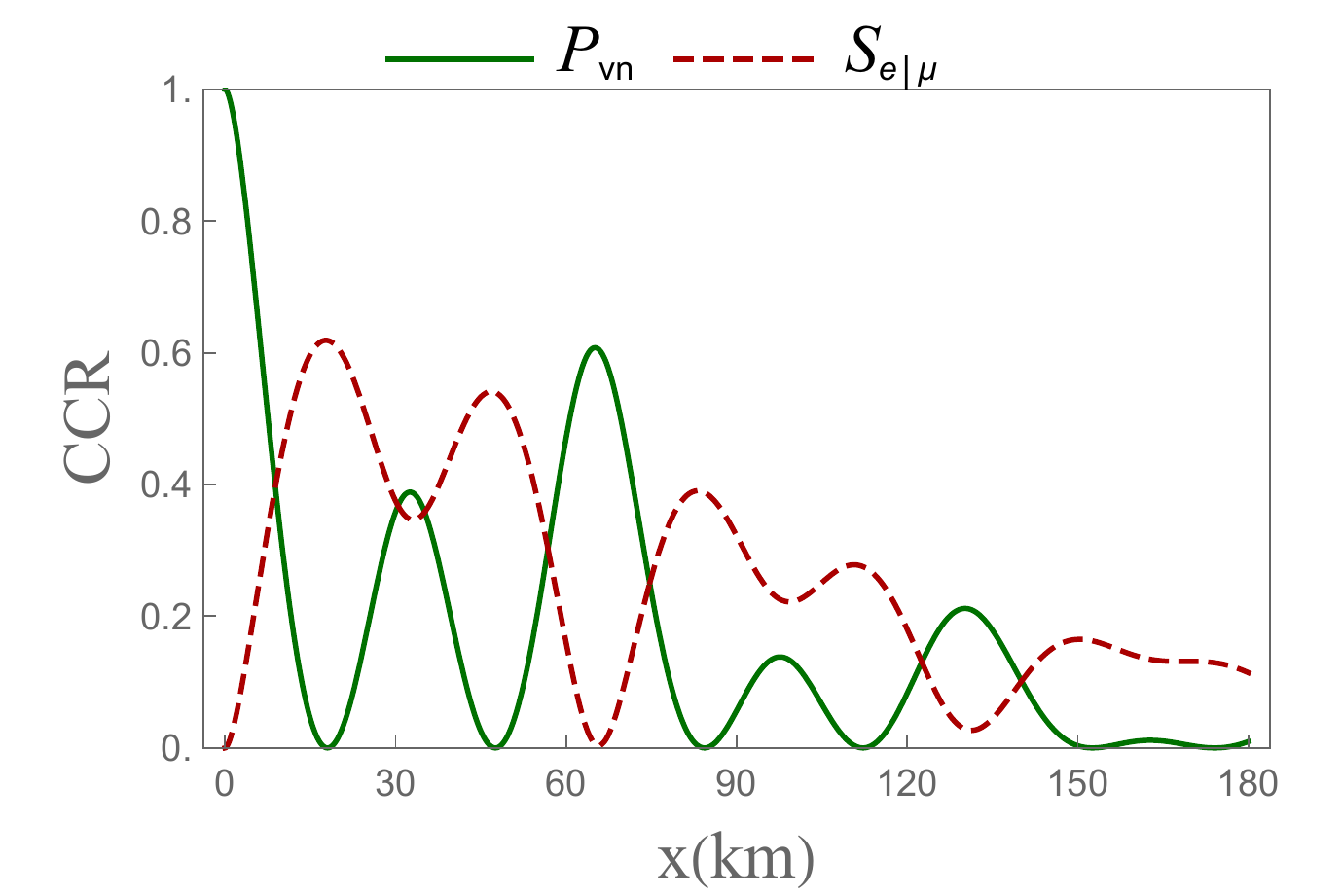}}}\\
 \subfloat[][\emph{MINOS}]{{\includegraphics[width =6 cm]{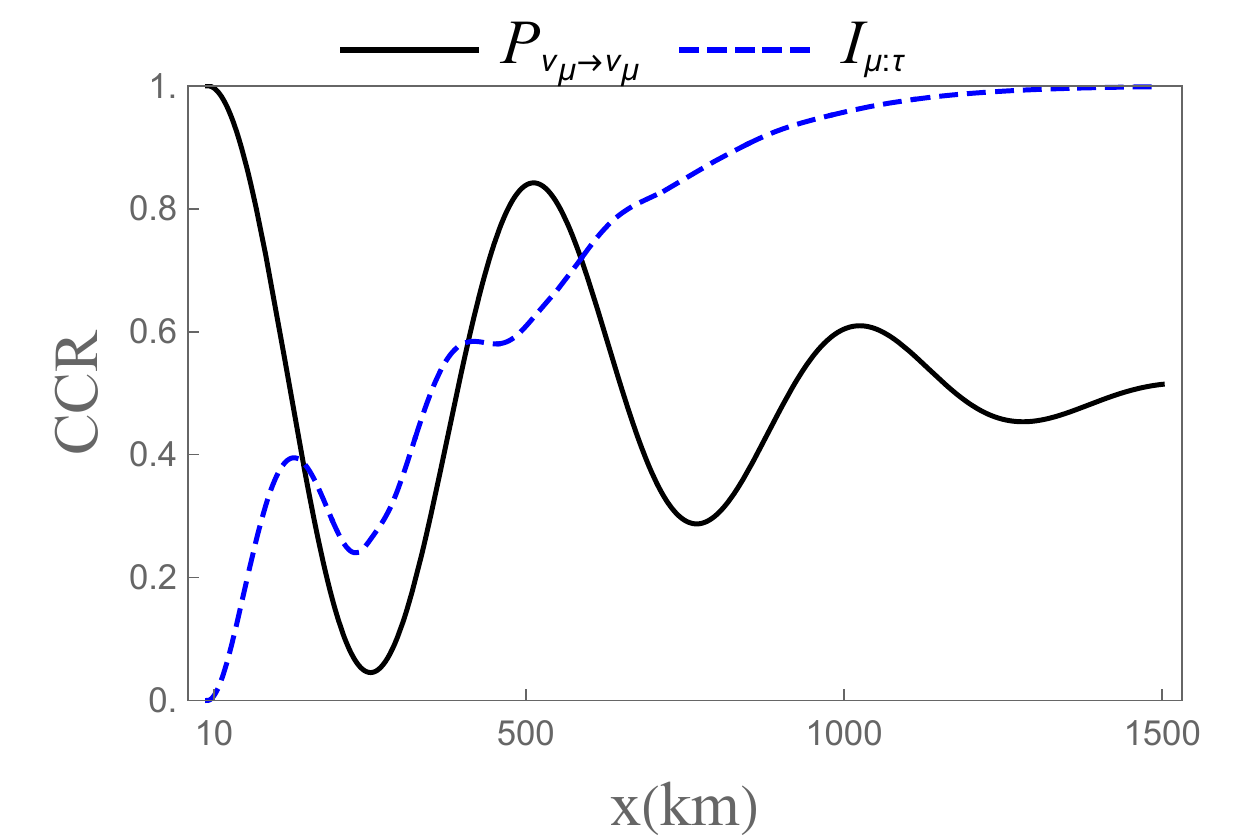}}\quad\quad
{\includegraphics[width =6.4 cm]{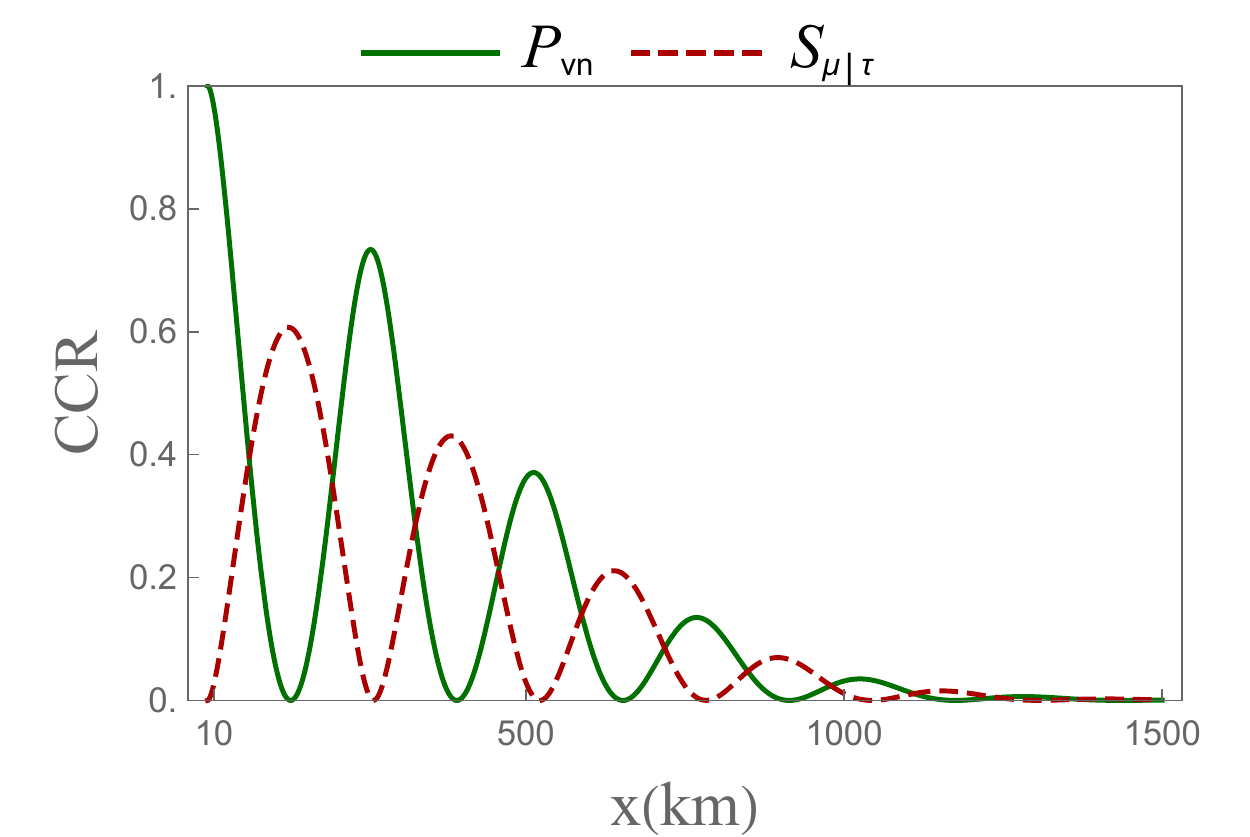}}}
\caption{On the left panels are shown the survival probability and the mutual information as a function of the propagation distance x(km) for a neutrino state \eqref{14}. On the right panels  are shown the conditional entropy and the predictability as a function of distance x(km). Parameters are given in Table 1.}
\label{fig1}
\end{figure}

In Fig.(\ref{fig1}) terms of Eq.(\ref{13}) are shown for the Daya Bay, Kamland and MINOS parameters, along with the survival probability, for comparison. We see that, in the case of the wave-packet approach, the terms included in the CCR, and their internal balancing, show non-trivial characteristics.  In fact, different values of the mixing angle associated to the three experiments lead to very different behaviors, especially in the asymptotic range. 

In the KamLand and Minos experiments, associated to higher values of the mixing angle, the mutual information\footnote{Note that for these two experiments, the mutual information asymptotically coincides with the quantum discord.} grows almost monotonically, \new{“engulfing” the other two terms, and keeping a high value even after oscillations are washed out. Due to the low value of the mixing angle, this aspect is not present for the Daya Bay parameters.} Furthermore, by looking to the left panels for KamLand and Minos experiments (high mixing angle values), it is very difficult to recognize in the mutual information \new{a behaviour exclusively dependent on the oscillation probability}.\new{ In fact, in those cases one can show that the oscillations displayed by all the quantities have components at different spatial frequencies which, in their own, are different than the oscillation frequency of the survival probability. This is highlighted for the Kamland parameters, for which $S_{e \vert \mu}$ exhibits a spatial frequency beating.} In this case, the intermediate value of the mixing angle gives rise to a not monotonically decreasing of the oscillations, with the presence of partial revivals.

\section{Conclusions}

In this Letter, we have applied the recently formulated complete complementarity relations (CCR) to neutrino flavor oscillations,  allowing us to completely characterize the quantumness of this phenomenon. We find that the quantum nature of mixed neutrinos contains features that go beyond the  flavor oscillations. Since the CCR formalism is complete, it can be connected to various quantitifiers previously studied in connection with this system, such as entanglement and NAQC. Our results confirms previous findings, showing that even after the complete spatial separation of the wave packets composing a flavor state, quantum correlations still persist \cite{Us}. 

We have considered the values of parameters associated to three experiments (Daya Bay, MINOS and KAMLAND), and  found that the long-distance behavior of quantum correlations is strongly dependent on the value of the mixing angle, which differs for the three cases. It remains an open question how such a quantumness associated to neutrino mixing and oscillations could be exploited as a (quantum) resource.

As an extension of our work, we plan to consider the instance of three flavor mixing and oscillations: in this case, any bipartition of the system will exhibit an internal structure which we expect could give rise to a non vanishing local coherence term, which is absent in the two flavor case. Furthermore, the extension to a quantum field description of neutrino flavor oscillations \cite{BV95,BV1} could reveal novel features of the phenomena discussed here.

\end{document}